\documentstyle[prl,aps]{revtex}
\input{epsf.tex}
\begin{document}
%\tighten
\wideabs{
%\twocolumn[{
%\narrowtext
%\draft
%\widetext
\title{Dissipative Landau--Zener Tunneling at Marginal Coupling}
\author{A. V. Shytov}
\address{L. D. Landau Institute for Theoretical Physics, 
2 Kosygina str., Moscow 117334, Russia}
%\date{\today}
\maketitle
%\mediumtext
\begin{abstract}
The Landau--Zener transition in a two level system can be suppressed or
enhanced by coupling to an environment, depending on the temperature
and the environment spectral function. 
We consider the marginal spectral function, when the 
dissipation effects are important for arbitrarily slow motion. 
Landau--Zener transition rate
demonstrates a non-trivial dependence of the on the ``bias'', i. e., 
on the rate of the two energy levels relative motion. 
The Landau--Zener transition is 
fully suppressed for the values of the bias below a 
threshold bias set by the coupling strength. 
Above the threshold, the transition rate 
for zero temperature is found 
using the instanton method. 
At finite temperature, the Landau--Zener transition rate
has a non-monotonic dependence on the coupling strength, being  
suppressed at the strong coupling. 
\end{abstract}
}
%]
%}
%\narrowtext
\noindent
{\it Introduction.}
The Landau--Zener problem\cite{LandauZener} 
deals with non-adiabatic transition in a two-state system 
under external bias. Physical examples include
current states in small metallic loops\cite{Gefen}, 
spin tunneling in magnetic molecules\cite{MolecularMagnets}, 
Andreev states in SNS junctions \cite{AndreevStates}, 
slow atomic and molecular collisions\cite{Atomic}, 
electron transfer in biomolecules\cite{Biophysics}. 
%It also describes the adiabatic flip of a qubit
%and thus may be important for the physics of quantum computing.
Real-world two-state systems are typically coupled to the environment,
and this coupling may lead to different physical effects.
This coupling provides dissipation and slows the tunneling down.
On the other hand, thermal noise in the environment may 
cause non-adiabatic transition.
The competition between these two effects makes the problem
of dissipative Landau--Zener tunneling very interesting.

Also, the Landau--Zener tunneling 
is of interest in the context of quantum computing, 
because
it  describes the adiabatic flip of a qubit.
During the process of the flip, 
the qubit is 
in a superposition of the pure ``on'' and ``off'' states  
%not in stable  ``on'' or ``off'' state, 
and thus may be very sensitive to environmental noise. Thus, the problem
of dissipative Landau--Zener tunneling may be important for the 
physical implementation of the qubit. 

Previous works on the 
dissipative Landau--Zener tunneling 
treated the problem either perturbatively or phenomenologically 
\cite{Gefen,GefenAdiabatic,Ultrahigh,Rammer,Shimshoni}.
Perturbative treatment, however, is insufficient
for adiabatic limit, because it
poses too strong limitations on coupling strength.
Phenomenological
approach is satisfactory only for high temperatures.
Also, previous works focused primarily on the case of Ohmic
coupling which is critical for equilibrium  tunneling\cite{CaldeiraLeggett}.
We will argue that the dissipation is relevant for Landau--Zener problem
when the coupling constant $\alpha$
has the same dimension as the velocity of levels relative motion $\nu$.
(We call the latter bias in the following.)
We call this coupling {\it marginal}.
(A physical example of such coupling is an SNS junction with two
Andreev states\cite{AndreevStates} connected to resistor.)
It turns out that the tunneling is blocked 
if the  bias is small: $\nu < \alpha$.
For larger bias, we derive an instanton solution describing dissipative 
Landau--Zener tunneling at low temperatures and find 
the tunneling probability:
\begin{equation}
\label{main-result}
 w \sim \exp \left( - \frac{\pi \Delta^2}{\nu - \alpha}\right)\ ,  
\end{equation}
where $\Delta$ is the energy gap, $\nu$ is the bias, and $\alpha$
is the coupling strength.
In the high temperature limit ($T \gg \hbar \Delta / 2 \nu$)
we derive master equation which takes into account both thermal
noise and dissipation. This equation solves the problem for all
temperatures larger than the crossover temperature, 
and can also incorporate non-equilibrium noise.

\noindent
{\it Model.}
It is conventional to represent the two states of Landau--Zener model 
using the  spin 1/2 basis. The Hamiltonian is
\begin{equation}
\label{hamiltonian-form}
\hat{H} = \nu t \hat{\sigma}_z + \Delta \hat{\sigma}_x 
+ \hat{H}_c + \hat{H}_{env}\ ,
\end{equation}
where $\nu$ is the bias, $\Delta$ is the energy gap,
and $\sigma_{x,z}$ are Pauli matrices.
The term $\hat{H}_c$ describes the coupling to the environment with
Hamiltonian $\hat{H}_{env}$.

If the coupling $\hat{H}_c$ is negligible, the model can
be solved exactly \cite{LandauZener}, and the non-adiabatic transition
probability is $w = \exp ( - \pi \Delta^2 / \nu)$.
In the adiabatic limit, $\nu \ll \Delta^2$,
the ``adiabatically frozen'' eigenvalues are:
\begin{equation}
\epsilon_{\pm}(t) = \pm \sqrt{(\nu t)^2 + \Delta^2}\ .
\end{equation}
The characteristic time of tunneling is $\tau_0 = \Delta / \nu$.

Following \cite{CaldeiraLeggett},
we model the environment by a set of oscillators
with coordinates $\hat{x}_i$ and momenta $\hat{p}_i$: 
\begin{equation}
\hat{H}_{env} = \sum_{i} \left(\frac{\hat{p}_{i}^2}{2} 
                + \frac{\omega_i^2 \hat{x}_i^2}{2}\right)
\ ,
\end{equation}
and couple the environment to the spin linearly:
\begin{equation}
\hat{H}_c = \hat{\sigma}_z \hat{X}(t)\ ; \quad 
\hat{X}(t) = \sum_{i} \gamma_i \hat{x}_i (t)\ . 
\end{equation}
We consider here only the diagonal coupling 
which is the main source of dephasing. 
The effect of environment depends only on the  spectral function:
\begin{equation}
J(\omega) = \sum_i \frac{\gamma_i^2}{2\omega_i}\, \delta (\omega - \omega_i)\ . 
\end{equation}
Without the loss of generality, in the limit of small $\omega$
one can consider power-like spectral functions: 
$J(\omega) = \alpha_s \omega^s$.
To estimate the effect of dissipation, we look at
the dimensionless ratio $\eta = \alpha_s / (\nu\tau_0^{s+1})$,
which gives the measure of  the dissipation 
with respect to the bias $\nu$.
In the adiabatic limit, when $\tau_0$ is large,
the dissipation is weak ($\eta \ll 1$) for $s > -1$,
and strong otherwise.
The marginal situation arising when $J = \alpha / \omega$
is a subject of this work.

Marginal coupling describes, e. g., 
single--channel SNS junction \cite{AndreevStates}
connected in series to the resistor.
There are two Andreev states in the junction.
%If the phase difference $\phi (t)$ across the junction is close to $\pi$, 
The dynamics of these states is governed by the Hamiltonian
\begin{equation}
\hat{H}_{SNS} = \frac{\Delta_0} \cos\frac{\phi(t)}{2} \, \sigma_z 
+ \Delta_0 \sqrt{1 - \tau} \sin\frac{\phi(t)}{2} \sigma_x\ , 
\end{equation}
where $\phi(t)$ is the superconducting phase difference
across the junction,
$\Delta_0$ is the superconducting gap in the leads, and $\tau$ is the
channel transmission coefficient.
Let $V$ is voltage drop across the whole circuit, and $V_R(t)$
is the voltage drop across the resistor.
Then, the phase difference is:
\begin{equation} 
\phi (t) = \frac{2 eVt}{\hbar} - \frac{2 e}{\hbar} \int \, V_R(t)\, dt\ . 
\end{equation}
In the vicinity of the point $\phi = \pi$, where the energy difference of
the two states is minimal, one may expand $H_{SNS}$ in $\phi$.
If there were no resistor, the system then would be described by Landau--Zener
model. 
To see what is the effect of the resistor, 
consider voltage fluctuations.
According to quantum Nyquist formula, the voltage $V_R$ fluctuates at
zero temperature as $\langle V_{R, \omega} V_{R, -\omega} \rangle
\sim R \omega$, where $R$ is the resistance, and $\omega$ is the 
frequency. 
Thus, phase fluctuations are:
\begin{equation}
\langle \delta\phi_{\omega}\, \delta\phi_{-\omega} \rangle 
\sim \frac{R}{\omega} \ . 
\end{equation}
Comparing this correlation function to $\langle X_\omega X_{-\omega} \rangle$, 
one may see that
the circuit is described by Landau--Zener model
with marginal dissipation. The parameters are identified as:
\begin{equation}
\nu = \frac{eV \Delta_0}{\hbar} \ ; 
\qquad 
\Delta = \Delta_0\sqrt{1 - \tau}\ ; 
\qquad
\alpha =  \frac{eR \Delta_0^2}{2\pi \hbar^2}\ . 
\end{equation}

There are two regimes of tunneling for different values of
environment temperature $T$. 
For $T \gg \hbar \Delta / \nu$ the tunneling is thermally assisted,
and the tunneling probability obeys Arrhenius law:
$w \sim \exp (- 2\Delta / T)$.
For low temperatures, the tunneling is due to non-adiabaticity, 
and the probability saturates at $T\to 0$. 
We consider these regimes separately.

\noindent
{\it Quantum regime.}
At low temperatures, we treat the nonequilibrium problem
by the Keldysh technique\cite{Keldysh,Schmid}. We introduce
two time contours describing evolution of wave function 
and its complex conjugate.
Standard contours going along the real time axis are 
inappropriate for the problem in question because the exponentially small 
adiabatic
transition probability results from a delicate destructive interference
of many oscillating contributions. 
We move Keldysh contours away from the real axis
to avoid oscillating terms.
Our choice of contours is similar to \cite{Ioselevich}, 
where it was used in the context of exciton autolocalization.
To illustrate the use of contours, 
consider the tunneling for $\alpha = 0$.
The transition probability is $w = \exp (-S_0)$,
with the action $S$ given by \cite{Landau-QM}:
\begin{equation}
\label{adiabatic-action}
S_0 = 2\mathop{\rm Im}\nolimits \int\limits_{t_1}^{i\tau_0} (\epsilon_{+}(t) -
\epsilon_{-}(t))\, dt = \frac{\pi \Delta^2}{\nu}\ .
\end{equation}
The integral is taken from any point $t_1$ on the real axis to $i\tau_0$,
where the adiabatic energy $\epsilon(t)$ has a square root singularity.
The two adiabatic states correspond to the branches 
$\epsilon_{\pm}(t)$, i.~e., to the two different sheets of the Riemann surface.

\begin{figure}
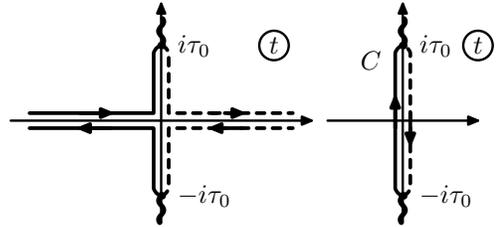

\begin{center}
\begin{tabular}{cc}
\epsfbox{fig1.eps} &
\epsfbox{fig2.eps}
\end{tabular}
\end{center}
\caption{
  (a) Keldysh contours. (b) Transformed contours.
  Dashed lines go on the other sheet of the Riemann surface.
  Wavy lines are branch cuts. 
}
\label{fig-1}
\end{figure}

The result (\ref{adiabatic-action})
may be represented using Keldysh contours
shown on Fig.\ref{fig-1}(a).
In the Keldysh formalism,
the forward contour always goes from $t = - \infty$ on the  real axis,
where the initial state is prepared,
to $t = +\infty$ where the final state is measured.
We draw the contour through the branching point
$t = i\tau_0$.
At this point the two energies $\epsilon_+(t)$
and $\epsilon_-(t)$ coincide, hence the adiabaticity is violated,
and the transition occurs.
It means that we change the branch $\epsilon_{-}(t)\to\epsilon_{+}(t)$,
and continue the contour on another sheet of Riemann surface.
Then, the contour
goes back to the real axis, and then to $t = +\infty$.
The backward  contour is a complex conjugate of
the forward contour.
In the adiabatic approximation, the segments parallel to the real axis give no
contribution to the imaginary part of the action,
and thus one may connect vertical segments into a single contour $C$
(see Fig.\ref{fig-1}(b)).
In this representation,
the action (\ref{adiabatic-action})
is given by
\begin{equation}
S_0 = \oint\limits_{C} \sqrt{(\nu t)^2 + \Delta^2}\, dt
\end{equation}

It turns out that the problem with $\alpha \neq 0$
has a solution of a similar structure:
there is a square root singularity at some point $t = i\tau_\alpha$
in the complex plane,
and the branch is changed when the contour passes this point.
In this case, the branching point $i\tau_\alpha$
has to be found self-consistently.
We derive the effective action on the contour $C$,
look for an instanton solution and find $\tau_\alpha$
from equations  of motion.
Let us introduce a field
$\Phi(t) = \nu t + X(t)$, so that the energy of the spin is 
$\sqrt{\Delta^2 + \Phi^2(t)}$.
Thus, the  two-level system action is
\begin{equation}
S_{TLS} = -i \oint_C dt\, \sqrt{\Delta^2 + \Phi^2(t)}\ . 
\end{equation}
(The Berry phase is neglected here since the ``magnetic field'' acting on the
spin is always in the $xz$--plane).
The dynamics of the environment
is described by $S_{env} + S_\lambda$, where
\begin{equation}
S_{env} = \frac{1}{2}\sum_{i} \left(\dot{x}_i^2 - \omega_i^2 x_i^2\right)\ ,
\end{equation}
and the term $S_\lambda$ enforces the constraint
$\dot{\Phi} - \dot{X} = \nu$: %, one may introduce
\begin{equation}
S_{\lambda} = -i \oint_C dt \lambda(t) \left(\dot{\Phi}(t) 
- \nu - \sum_{i}\gamma_i \dot{x}_i (t)\right)
\end{equation}
We integrate out $\Phi(t)$ and $x_i(t)$ in the saddle-point approximation
and find the effective action in terms of the Lagrange multiplier 
$\lambda(t)$:
\begin{eqnarray}
\label{action}
S_{eff} &=& i \Delta \oint dt\, \sqrt{1 - \dot{\lambda}^2} 
         +  i\nu \oint dt\, \lambda(t) 
            \nonumber \\
        &+& \frac{\alpha}{4\pi} \oint\oint dt\, dt' \,
\frac{(\lambda(t) - \lambda(t'))^2}{(t - t')^2}\ .
\end{eqnarray}

The solution describing Landau--Zener tunneling has different signs on
different sheets of the Riemann surface:
$\lambda (t-0) = - \lambda (t+0)$.
Taking a saddle point with respect to $\lambda(t)$, one arrives to
the equation of motion:
\begin{equation}
\label{saddle-point-equation}
  \Delta \frac{d}{dt}\frac{\dot{\lambda}}{\sqrt{1 - \dot{\lambda}^2}} 
 = - \nu + \frac{i \alpha}{\pi} 
   \int\limits_{-\tau_\alpha}^{\tau_\alpha} 
       \frac{\lambda(t')\, dt'}{(t - t')^2}
   \ . 
\end{equation}
For $\alpha = 0$ the solution is $\lambda_0(t) = \sqrt{\tau_0^2 + t^2}$.
Interestingly,
the integral in (\ref{saddle-point-equation}) with $\lambda = \lambda_0(t)$
is constant for $t \in [-i\tau_0; i\tau_0]$,
and the r.h.s. of (\ref{saddle-point-equation}) preserves its form.
Thus, one may look for the solution of the form
$\lambda(t) = \sqrt{\tau_\alpha^2 + t^2}$.
%One finds then from Eq. (\ref{saddle-point-equation}):
From Eq. (\ref{saddle-point-equation})
one finds $\tau_\alpha = \Delta / (\nu - \alpha)$.
Substituting it into (\ref{action}), one obtains the action 
\begin{equation}
\label{quantum-action}
S = \frac{\pi \Delta ^2}{\nu - \alpha}\ , 
\end{equation}
and the tunneling probability $\exp (-S)$ given by Eq. (\ref{main-result}).

This solution gives an instanton for $\nu > \alpha$.
Otherwise, for $\nu < \alpha$,
there is no saddle point solution and the tunneling
is impossible. This happens because the friction force
is larger than the bias,
and the energy put into the system by the bias source,
is fully dissipated into the environment.
Because of that, there is no level crossing, even at complex times.
The tunneling time $\tau_{\alpha}$ diverges at
$\nu = \alpha$. 
Because of that, the dynamics far from level crossing
point becomes essential. The linear approximation on which 
Eq.~(\ref{hamiltonian-form}) is based may break down. 
Also, for $\nu \simeq \alpha$, the tunneling time
may be comparable to the time $t_{\infty}$ when the system
is prepared and measured. Thus, the Landau--Zener problem makes
no sense for $\nu$ too close to $\alpha$. 
The behaviour of the system is determined by the details
of the dynamics far from the avoided crossing point.

In the perturbative regime, $\alpha \ll \nu$,  
the action (\ref{quantum-action}) can be expanded in $\alpha$:
\begin{equation}
S = S_0 + \frac{\pi \Delta^2 \alpha}{\nu^2}\ .
\end{equation}
This result contradicts to Ao et al.,\cite{Rammer}, 
who found an $\alpha$--independent
tunneling probability. We believe that the result\cite{Rammer}
is incorrect because of improper choice of shakeup force $\zeta(t)$ that does not
describe tunneling. 

The above solution shows that at $T = 0$ the environment effect is
purely dissipative, and the coupling to the environment reduces the
transition probability. This happens because zero-point fluctuations
of the environment %are zero-point motion, and
do not lead to
a real transition. At finite temperatures, the environment may
transfer the energy to the two-level system, thus increasing the
transition probability. If the temperature is low,
$T \ll T_0 = \hbar / \tau_\alpha$,
the transition still has mostly quantum nature.
In this limit, the finite temperature
can be taken into account by requiring the instanton to be periodic
in imaginary time, $\lambda(t) = \lambda(t +i \beta)$, 
with the period $\beta = 1/ T$.
Thus, one has to consider a periodic system of cuts,
$t \in [-i\tau + i \beta n, i\tau + i\beta n]$.
Note that the nonlocal term in (\ref{action}) gives rise to interaction
between instantons on different cuts.
If $\beta \gg \tau_\alpha$,
the interaction is weak and can be treated perturbatively.
The correction to the action per period is:
\begin{equation}
\delta S = - \frac{\pi \alpha \tau_\alpha^4}{2} \sum_{n \neq 0}\frac{1}{\beta^2n^2} 
\end{equation}
After computing the sum over $n$, one obtains:
\begin{equation}
\label{action-temperature}
S(T) = S(T=0)\left(1 - \frac{\pi^2}{6}\frac{\alpha T^2 \tau_\alpha^2}{(\nu -
\alpha)}\right)\ , 
\end{equation}
and the tunneling probability is $w \sim \exp (-S(T))$, as before.
Eq. (\ref{action-temperature}) is applicable when the correction
is small and thus does not lead to non--monotonous $T$--dependence.

\noindent
{\it Classical regime.}
%Now we analyze thermally assisted Landau--Zener tunneling.
If the temperature of the environment is high ($T \gg \hbar/\tau_\alpha$),
the tunneling is thermally assisted, and
the effect of environment can be divided into slow regular motion
with frequencies $\omega \sim 1/ \tau_\alpha$ and fast Langevin noise
with $\omega \gg 1/\tau_\alpha$.
(Only noise with $\omega > 2\Delta$ contributes to the 
transition probability, and this separation is valid in the adiabatic limit.)
Then the Hamiltonian is:
\begin{equation}
\hat{H} = F(t)\,\hat{\sigma}_z + \Delta \hat{\sigma}_x + \hat{u}(t) \sigma_z\ , 
\end{equation}
where $F(t) = \nu t + \langle X(t) \rangle$ is a regular part, and
$\hat{u}(t) = \hat{X}(t) - \langle X(t) \rangle$ is a fluctuating part.
Since the first two terms are slow functions of time,
one may consider them in the adiabatic approximation, and treat
the noise $\hat{u}(t)$ as a perturbation causing non-adiabatic
transition. This implies that the noise contribution to the transition
amplitude is larger than the non-adiabatic correction.
The eigenstates
of the frozen Hamiltonian are:
\begin{equation}
\psi_{+}(t) = 
\left( 
\begin{array}{c}
\cos\frac{\theta(t)}{2} \\
\sin\frac{\theta(t)}{2} 
\end{array}
\right) \ ; \quad
\psi_{-}(t) = 
\left( 
\begin{array}{r}
- \sin\frac{\theta(t)}{2} \\
  \cos\frac{\theta(t)}{2} 
\end{array}
\right) \ ,
\end{equation}
where $\tan \theta(t) = \Delta / F(t)$\ .
In this basis the Hamiltonian is:
\begin{equation}
\hat{H} = -\epsilon (t) \hat{\sigma}_z 
    + \hat{u}(t) (\cos\theta(t)\, \hat{\sigma}_z 
    + \sin\theta(t)\,\hat{\sigma}_x
                 )\ , 
\end{equation}
where $\epsilon(t) = \Delta / \sin\theta(t)$ is the adiabatic energy.
The perturbation theory with respect to $\hat{u}(t)$ 
gives the transition rates
%with excitation of $m$th state of the environment into $n$th:
\begin{equation}
\label{transition-rates-w}
w_{\pm, m\to n} = 
\left|\int\limits_{-\infty}^{\infty} e^{\pm i\lambda (t)}\, 
\sin\theta(t)\, \langle n  | \hat{u}(t)| m \rangle \, dt\right|^2
\ 
\end{equation}
between the states $|m\rangle$ and $|n\rangle$ of the environment. 
In Eq. (\ref{transition-rates-w}), $\dot{\lambda}(t) = \epsilon (t)$, 
and the subscript $(\pm)$ denotes initial spin state.
Tracing out the environment, one finds the Landau--Zener transition rates 
\begin{equation}
w_{\pm} = \int\limits_{-\infty}^{\infty} dt \int\limits_{-\infty}^{\infty} dt' \, 
e^{\pm i(\lambda(t) - \lambda(t'))}\, \sin\theta(t)\, \sin\theta(t')\, 
\langle \hat{u}(t) \hat{u}(t')\rangle\ . 
\end{equation}
This integral contains a fast oscillating function, and is determined by the
region where $|t - t'| \sim 1/\epsilon(t) \ll \tau_0$.
Therefore, one may approximate in the prefactor $t \simeq t'$,
integrate over $t-t'$ and find the Landau--Zener transition rates:
\begin{equation}
\label{transition-rate}
w_{\pm} = \int\limits_{-\infty}^{\infty} K(\omega = \pm 2\epsilon(t))\,
\sin^2\theta(t) \, dt \ .  
\end{equation}
Here 
\begin{equation}
\label{noise-spectrum}
K(\omega) =  \int\limits_{-\infty}^{\infty} e^{i\omega t} 
\langle \hat{u}(t) \hat{u}(0) \rangle\, dt  
\end{equation}
is the noise spectrum. In equilibrium, it can be expressed in terms of
the spectral function $J(\omega)$ and the Bose distribution
function $N(\omega)$: 
\begin{equation}
K(\omega) = J(|\omega|)(N(|\omega|) + \theta(\omega))\ . 
\end{equation}
The result (\ref{transition-rate}) is also true for non-equilibrium noise
with the proper choice of $K(\omega)$. 

To find the still unknown function $\theta(t)$, consider 
the force exerted by the rotating spin on the
environment, $f(t) = \langle \hat{\sigma}_z(t) \rangle = \pm\cos\theta(t)$.
The response to that force is, in Fourier representation,
$\langle X_\omega \rangle = -i J(\omega) f_\omega$.
Then, one may write an equation for $\theta(t)$:
\begin{equation}
\dot{F}(t) = \nu + \langle\dot{X}(t)\rangle = \nu \mp \alpha \cos\theta(t)\ . 
\end{equation}
Since $F(t) = \Delta / \tan\theta(t)$, one has:
\begin{equation}
\Delta \dot{\theta} = \sin^2\theta(t) (\nu \mp \alpha \cos\theta (t))\ .  
\end{equation}
For small bias, $\nu < \alpha$, there is no Landau--Zener tunneling,
and situation is similar to quantum limit. 
If the bias $\nu > \alpha$, the transition probability is: 
\begin{equation}
\label{transition-rate-theta}
w_{\pm} = \int\limits_{0}^{\pi} r_{\pm}(\theta)\, d\theta \ ; \quad
r_{\pm}(\theta) = 
\frac{\Delta K(\pm \frac{2\Delta}{\sin\theta})}{\nu \mp \alpha\cos\theta} \, 
\ . 
\end{equation}
Note that the small bias $\nu$ appears in the denominator.
Because of that, the probability computed from (\ref{transition-rate-theta})
can be quite large, and the perturbation theory on which this equation
is based, appears to break down. This happens because the motion at small
bias is slow, and a small noise acts on the system 
for a long time, giving rise to a large tunneling probability.

To solve the problem, note that
according to (\ref{transition-rate}), there is no interference
between transition amplitudes at different times $t$.
This happens because the transitions are due to random noise,
and the amplitudes of separate transitions have random phase.
In this situation, one may use master equation,
with the transition rate per unit time from (\ref{transition-rate}).
Since $\theta(t)$ depends on the history of the system,
it is convenient to use
$\theta$ as an independent variable instead of $t$.
Master equation, written in the variable $\theta$,  has the form:
\begin{equation}
\label{master-equation}
\frac{d{P}_{+}(\theta)}{d\theta} 
= - \frac{d{P}_{-}(\theta)}{d\theta} 
=  -r_{+}(\theta) P_{+}(\theta) + r_{-}P_{-}(\theta)
\ ,
\end{equation}
where $P_{\pm}(\theta)$ are occupancies of two adiabatic states.
The initial condition is $P_{+}(0) = 0$, $P_-(0) = 1$. 

Eq. (\ref{master-equation}) can be solved in a general form.
However, for simplicity, we consider
the two cases: (i) $T \ll \Delta$ and (ii) $T \gg \Delta$.
For $T \gg \Delta$, the transition rates $r_+$ and $r_-$ are equal.
The noise correlator is $K(\omega) = \alpha T / \omega^2$.
The calculation gives for $w = P_{+}(\theta = \pi)$:
\begin{equation}
\label{transition-rate-ultrahigh}
w = \frac{1}{2}\,(1 - e^{-2R})\ ; \quad 
R = \int\limits_{0}^{\pi} r_{-}(\theta)\, d\theta\ .
\end{equation}
Computing the integral, one obtains:
\begin{equation}
R =  \frac{\pi T}{4\Delta} 
\left(\frac{\nu}{\alpha} - \sqrt{\frac{\nu^2}{\alpha^2} - 1}\right)\ . 
\end{equation}
In the limit $T\to \infty$,  Eq. (\ref{transition-rate-ultrahigh})
predicts $w\to 1/2$,
the result of a phenomenological approach.
Note that (\ref{transition-rate-ultrahigh}) is true only
when $ R \ll \Delta^2/\nu$, otherwise non-adiabaticity is 
again important, and the exponential correction
has different structure\cite{Ultrahigh,Rammer}. 

We  consider now the case of intermediate temperatures:
$ \hbar / \tau_\alpha \ll T \ll \Delta$.
The transition $|\psi_-\rangle\to|\psi_+\rangle$
requires energy transfer $2\Delta$,
while the inverse transition does not,
and the transition rates $r_{+}$ and $r_{-}$ are not equal.
Therefore, the transition from the lower state to the upper one
occurs at $\theta = \pi / 2$, when the energy difference is minimal.
This transition can be considered as an instant excitation
occurring at $\theta \simeq \pi /2$ during the time
$t_{ex} \sim \sqrt{\Delta T} / \nu$.
It is followed by a decay from the upper state.
Thus, the transition probability is the product of the probability
of excitation  at $\theta = \pi / 2$ and the probability that the
excited state survives for $\pi /2 < \theta < \pi $.
From the master equation (\ref{master-equation}) one finds 
the excitation probability:
\begin{equation}
P_{ex} = \int\limits_{0}^{\pi} r_{-}(\theta) d\theta = 
\frac{\alpha}{\nu} \sqrt{\frac{\pi T}{\Delta}} e^{- 2\Delta/ T}
\ , 
\end{equation}
and the survival probability:
\begin{equation}
P_s = \exp\left( - \int\limits_{\pi/2}^{\pi} r_{+}(\theta) \, d\theta\right) = 
\sqrt{
1 - \frac{\alpha}{\nu}
}
\ . 
\end{equation}
Finally,  the transition probability is:
\begin{equation}
w = P_{ex} P_s = \frac{\alpha}{\nu} 
\sqrt{
1 - \frac{\alpha}{\nu}
%\right)
}
\sqrt{\frac{\pi T}{\Delta}} e^{-2\Delta/T} \ . 
\end{equation}
Note that the dependence of this expression on $\alpha$ is non-monotonous.
This is due to the fact that the coupling to the environment provides both
external noise and dissipation.
The transition probability reaches the maximal value at 
$\alpha = 2 \nu / 3$.

\noindent
{\it Conclusion.} 
We studied dissipative Landau--Zener tunneling
marginally coupled to environment in both quantum and classical
limits. At $T = 0$ the effect of environment is purely dissipative,
whereas at high temperatures thermal fluctuations in the
environment increase the transition probability, leading to
non--monotonous dependence of the transition rates on
%and the tunneling probability depends non-monotonously on 
the coupling strength $\alpha$.
At $\nu < \alpha$ the tunneling is blocked. 
Thus, the interplay between fluctuations and dissipation
is manifest in the marginal coupling model.

\noindent {\it Acknowledgements}.
I am grateful to M.~V.~Feigel'man, A.~S.~Ioselevich, L.~S.Levitov, 
D.~Esteve, and C.~Urbina for stimulating and illuminating discussions.
This research was supported by the Russian Ministry of Science via the 
program ``Physics of quantum computing'', and by
RFBR grant 98-02-19252.

\end{document}